# Visualizing higher-fold topology in chiral crystals

Tyler A Cochran*,†,[1] Ilya Belopolski*,[1] Kaustuv Manna,[2,3] Mohammad Yahyavi,[4] Yiyuan Liu,[5] Daniel S. Sanchez,[1] Zi-Jia Cheng,[1] Xian P. Yang,[1] Daniel Multer,[1] Jia-Xin Yin,[1] Horst Borrmann,[2] Alla Chikina,[6] Jonas A. Krieger,[6,7] Jaime Sánchez-Barriga,[8] Patrick Le Fèvre,[9] François Bertran,[9] Vladimir N. Strocov,[6] Jonathan D. Denlinger,[10] Tay-Rong Chang,[4] Shuang Jia,[5] Claudia Felser,[2] Hsin Lin,[11] Guoqing Chang†,[12] and M. Zahid Hasan†[1,13,14]

[1] *Laboratory for Topological Quantum Matter and Advanced Spectroscopy (B7), Department of Physics, Princeton University, Princeton, New Jersey 08544, USA*
[2] *Max Planck Institute for Chemical Physics of Solids, Dresden, Germany*
[3] *Department of Physics, Indian Institute of Technology Delhi, Hauz Khas, New Delhi 110016, India*
[4] *Department of Physics, National Cheng Kung University, Tainan 70101, Taiwan*
[5] *International Center for Quantum Materials, School of Physics, Peking University, Beijing, China*
[6] *Swiss Light Source, Paul Scherrer Institute, Villigen, Switzerland*
[7] *Laboratory for Muon Spin Spectroscopy, Paul Scherrer Institute, Villigen, Switzerland*
[8] *Helmholtz-Zentrum Berlin für Materialien und Energie, Elektronenspeicherring BESSY II, Albert-Einstein Strasse 15, 12489 Berlin, Germany*
[9] *Synchrotron SOLEIL, L'Orme des Merisiers, Saint-Aubin-BP 48, 91192 Gif-sur-Yvette, France*
[10] *Advanced Light Source, Lawrence Berkeley National Laboratory, Berkeley, California 94720, USA*
[11] *Institute of Physics, Academia Sinica, Taipei 11529, Taiwan*
[12] *Division of Physics and Applied Physics, School of Physical and Mathematical Sciences, Nanyang Technological University, 21 Nanyang Link, 637371, Singapore*
[13] *Princeton Institute for Science and Technology of Materials, Princeton University, Princeton, New Jersey 08544, USA*
[14] *Lawrence Berkeley National Laboratory, Berkelely, California 94720, USA*

Novel topological phases of matter are fruitful platforms for the discovery of unconventional electromagnetic phenomena. Higher-fold topology is one example, where the low-energy description goes beyond Standard Model analogs. Despite intensive experimental studies, conclusive evidence remains elusive for the *multi-gap topological nature of higher-fold chiral fermions*. In this Letter, we leverage a combination of fine-tuned chemical engineering and photoemission spectroscopy with photon energy contrast to discover the higher-fold topology of a chiral crystal. We identify all bulk branches of a higher-fold chiral fermion for the first time, critically important for allowing us to explore unique Fermi arc surface states in multiple inter-band gaps, which exhibit an emergent ladder structure. Through designer chemical gating of the samples in combination with our measurements, we uncover an unprecedented multi-gap bulk boundary correspondence. Our demonstration of multi-gap electronic topology will propel future research on unconventional topological responses.

Traditional topological quantum states have often been characterized by a single Chern number, such as the chiral charge of a Weyl fermion or the quantum anomalous Hall (QAH) conductance of a two-dimensional insulator [1–6]. The observation and manipulation of multi-Chern number topological structures promises to open new opportunities for fundamental physics and applications [7, 8]. Already, sequences of Chern insulating states have been observed in magic-angle twisted bilayer graphene [9–15], while large Chern numbers were engineered by the assembly of multiple QAH layers [2, 16]. In both cases the effects arise only at $\sim$10 meV energy scales and $\sim$200 mK temperatures, limiting the ability for spectroscopic study and technological application. However, multi-Chern number states are also associated with higher-fold chiral fermions in three-dimensional bulk topological semimetals, which can naturally possess a 1 eV energy scale at room temperature [6]. This prospect, if achieved, will allow for the study of generalizations of the quantized circular photogalvanic effect [17–20], exotic multi-Fermi arc nonlinear optical effects [21, 22], and multi-Fermi arc quantum oscillations [23]. However, despite this promising outlook, the multi-Chern nature of higher-fold chiral fermions has proven challenging to observe, despite considerable effort on stoichiometric B20 compounds [24–27]. Therefore, an urgent experimental challenge is to obtain a degree of tunability in these materials in order to understand their unconventional topological nature.

Higher-fold chiral fermions can be understood as a generalization of more familiar Weyl fermions, being themselves two-fold chiral fermions. For Weyl fermions, one topological invariant is defined within the band gap on $k$-space manifolds enclosing a two-fold degeneracy, the chiral charge $C$ [Fig. 1(a)]. As a consequence, one set of boundary states is topologically protected within the gap [Fig. 1(b)]. By extending this paradigm to higher-fold fermions, the set of possibilities becomes more diverse. Dirac-like nonchiral fermions have a chiral charge of zero in each gap [28–32], whereas Weyl-like chiral fermions have nonzero chiral charge in multiple gaps [18, 33–36]. For an $N$-fold chiral fermion, we introduce the multi-gap chiral charge $\mathbf{C} = (C_1, C_2, ..., C_{N-1})$, which includes one integer for each band gap [Fig. 1(c)]. From here, the bulk-boundary correspondence specifies that in gap $i$ there are $C_i$ chiral surface states (Fermi arcs). These states are also called chiral because of the presence of a net nonzero number of left/right moving quasi-particles for a chosen chemical potential along a closed path in the surface Brillouin zone. In the multi-gap case, there are a net nonzero number of right/left movers in multiple gaps, leading to chiral Fermi arcs that are stacked in the energy direction [Fig. 1(d)]. Therefore, to discover an $N$-fold chiral fermion with higher-fold topology, one needs to directly observe all the bulk branches [Fig. 1(c)], and use the Fermi arcs on the surface to determine the multi-gap chiral charge [Fig. 1(d)].

In order to establish the topological nature of any band structure, the bulk-boundary correspondence must be rigorously established experimentally [37–39]. In the case of higher-fold chiral fermions, the two key experimental signatures are: (1) resolving all the bulk bands that become degenerate at a higher-fold chiral fermion, and (2) showing nontrivial Chern number in each interband gap. Verifying these criteria amounts to confirming the multi-gap bulk boundary correspondence. In higher-fold topological material candidates to date, giant Fermi arc states and linear dispersions have been observed, indicating the presence of Weyl-like quasiparticles [24–27]. However, neither criterion for higher-fold topology listed above has been met by any experiment, on any material [40].

In the search for an ideal material candidate to study higher-fold topology, we consider crystals in structurally chiral space group $P2_13$ (#198), where non-zero Chern numbers and bulk cone-like dispersions have recently been observed [24–27]. Materials in this space group are promising because conical bands are predicted to arise from a three-fold chiral fermion at the Γ point, naturally providing a platform for multiple topological band gaps [Figs. 1(e), 1(f), and 1(g)] [33, 34].

In this Letter, we chemically engineer the substitutional alloy $Rh_{1-x}Ni_xSi$, to realize tunable chemical gating [41–45]. Rigorous Laue, as well as single crystal, X-ray diffraction measurements show that our sample exhibits excellent crystal quality, and possesses the desired structurally chiral space group $P2_13$ (#198). Detailed chemical analysis indicates that our $Rh_{1-x}Ni_xSi$ sample possess $x = 0.05$. The Flack parameter was refined to -0.01(11) throughout the sample, indicating that only one structurally chiral domain is present. Importantly, the presence of Ni does not induce any magnetism as the sample was found to be diamagnetic down to 10 K.

Using bulk-sensitive soft X-ray angle-resolved photoemission spectroscopy (SXARPES), we first search for higher-fold degeneracies using 550 eV incident photons [46–50]. We find that for the M-Γ-M binding energy vs crystal momentum ($E_B$ vs $k$) dispersion, three bands become degenerate 240 meV below the Fermi level at the Γ point, which is also reflected in the *ab initio* calculation [Fig. 2(a)]. Away from the Γ point the energy separation between all three bands becomes clear; we define gaps 1 and 2 to be the separations between bands 1, 2 and 3 away from the Γ point. To further verify our finding of a three-fold degeneracy, we measure the $E_B$ vs $k$ dispersion along another path, X-Γ-X [Fig. 2(b)]. Here we also observe a three-fold degeneracy at the Γ point. Along this new path, gap 1 is diminished compared to M-Γ-M. We confirm our results by systematically examining $E_B$ vs $k$ dispersions slightly away from the Γ point. As the X-Γ-X dispersion is shifted in the $k_y$ direction, band 3 quickly retreats above the Fermi level, while band 2 disperses very little, and band 1 falls to deeper binding energies [Figs. 2(c) and 2(d)]. Our experimental results confirm the bulk dispersion of a three-fold chiral fermion: a linear crossing between bands 1 and 3 is degenerate with a nearly flat band 2 at the Γ point. These three bands naturally give rise to two gaps in which the topological surface states should reside.

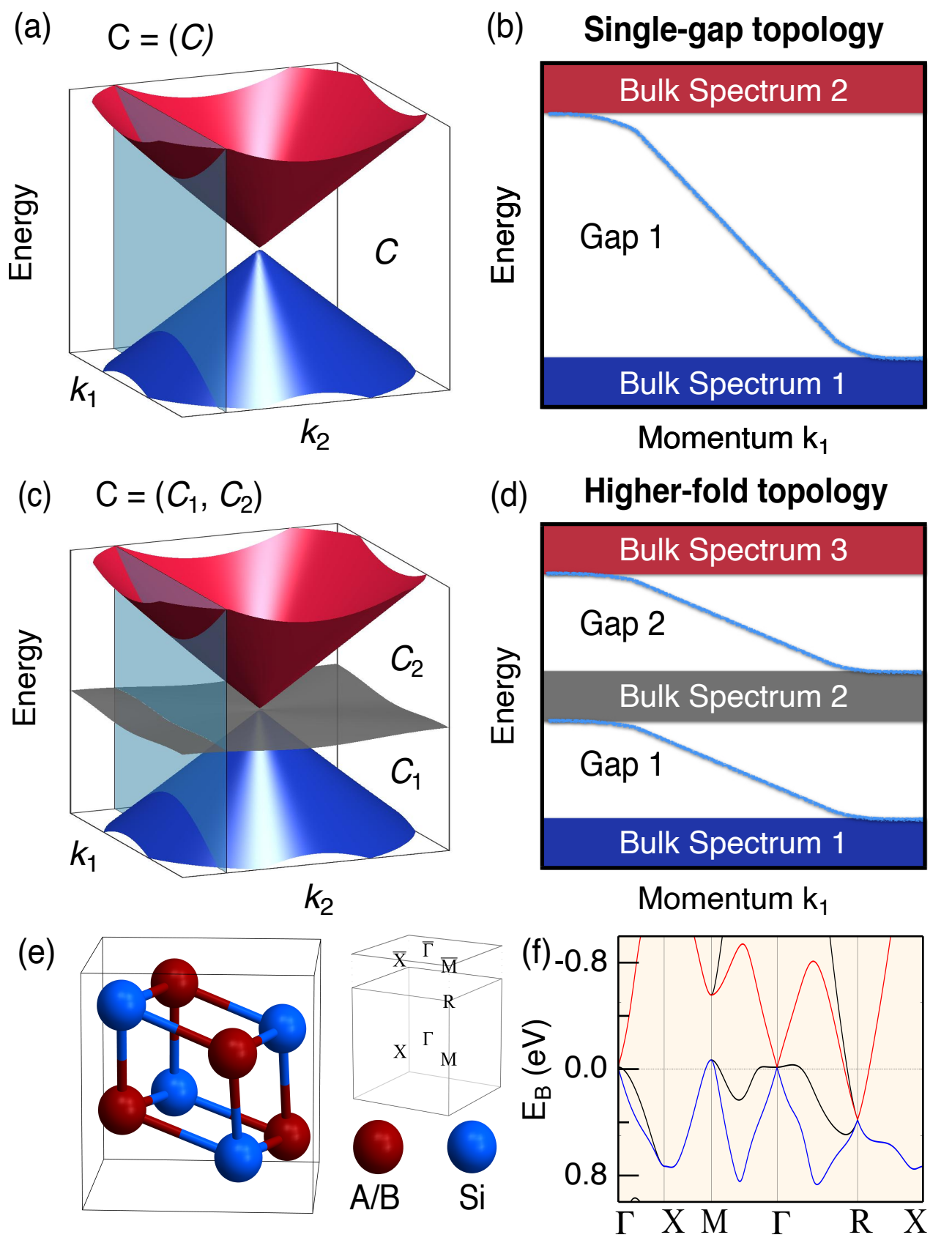

FIG. 1. Higher-fold topology beyond Weyl and Dirac. (a) Schematic of a two-band system corresponding to a Weyl fermion in three dimensions. (b) Schematic of bulk-boundary correspondence of a one-gap system with a boundary chiral mode. The momentum path corresponds to the blue plane in Fig. 1(a). (c) Schematic of a chiral fermion with three bulk branches and two topologically nontrivial energy gaps. (d) Schematic of bulk-boundary correspondence of a two-gap system with boundary chiral modes. The momentum path corresponds to the blue plane in Fig. 1(c). (e) Crystal structure of the $A_{1-x}B_xSi$ substitutional alloy in the $P2_13$ space-group (#198), with A=Rh and B=Ni. The Bulk and (001) surface Brillouin zone. (f) *Ab initio* band structure calculation along high-symmetry lines of RhSi.

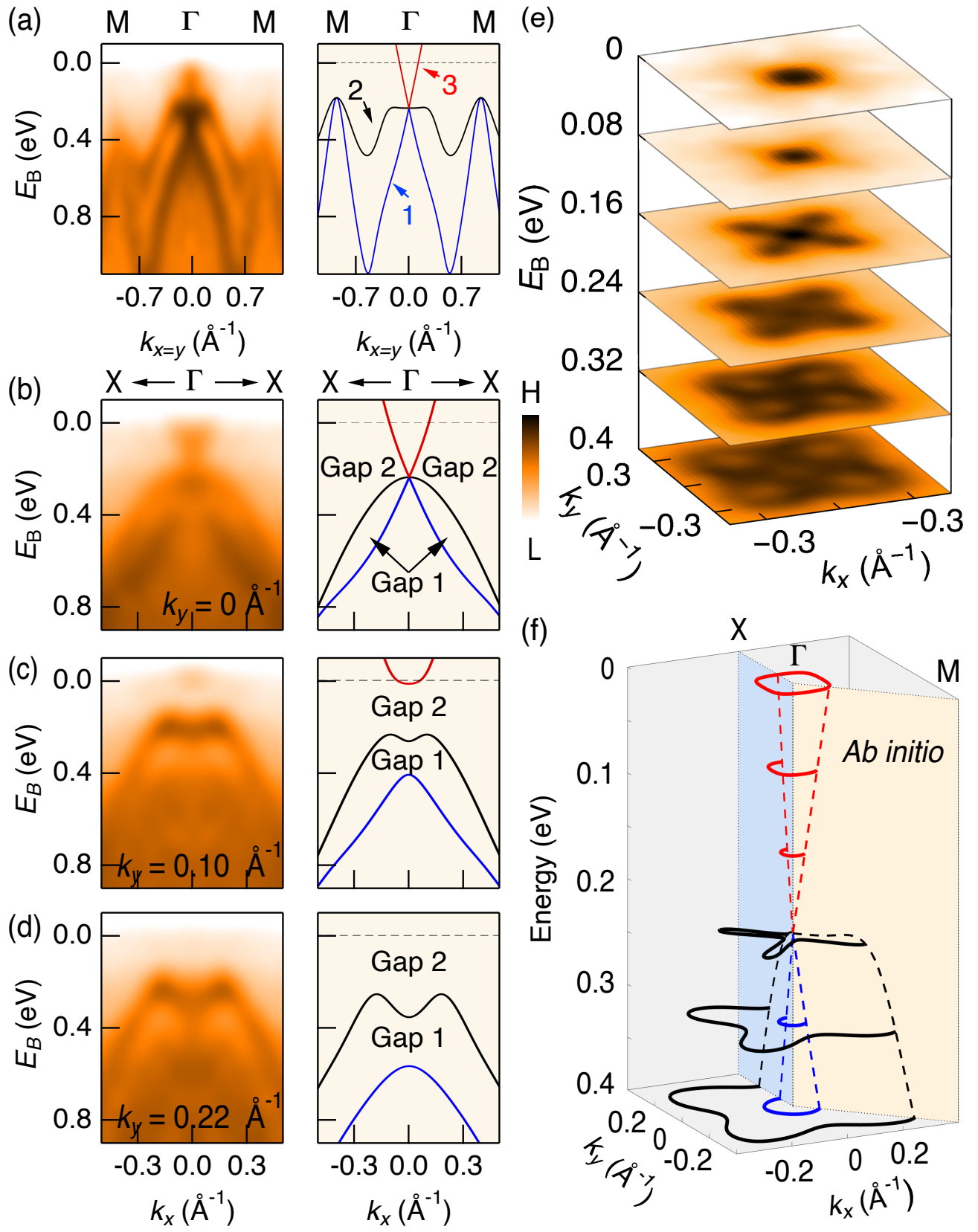

FIG. 2. Topological chiral fermion at the Γ point. (a) Bulk sensitive SXARPES measured using 550 eV incident photons and *ab initio* $E_\mathrm{B}$ vs $k$ spectrum along the M-Γ-M line, showing three bands becoming degenerate at the Γ point. (b) SXARPES and *ab initio* spectra along the X-Γ-X line, confirming three bands dispersing away from the degeneracy point. (c) SXARPES $E_\mathrm{B}$ vs $k$ spectrum and *ab initio* calculation along a path shifted to $k_y = 0.1$ Å$^{-1}$. Along this momentum path, gap 1 and gap 2 can be identified. (d) SXARPES spectrum and *ab initio* calculation along a path further shifted to $k_y = 0.22$ Å$^{-1}$. (e) SXARPES measured stack of constant $E_\mathrm{B}$ contours at the Γ point, showing the band structure evolution through the degeneracy point. (f) *Ab initio* calculation of constant $E_\mathrm{B}$ stack.

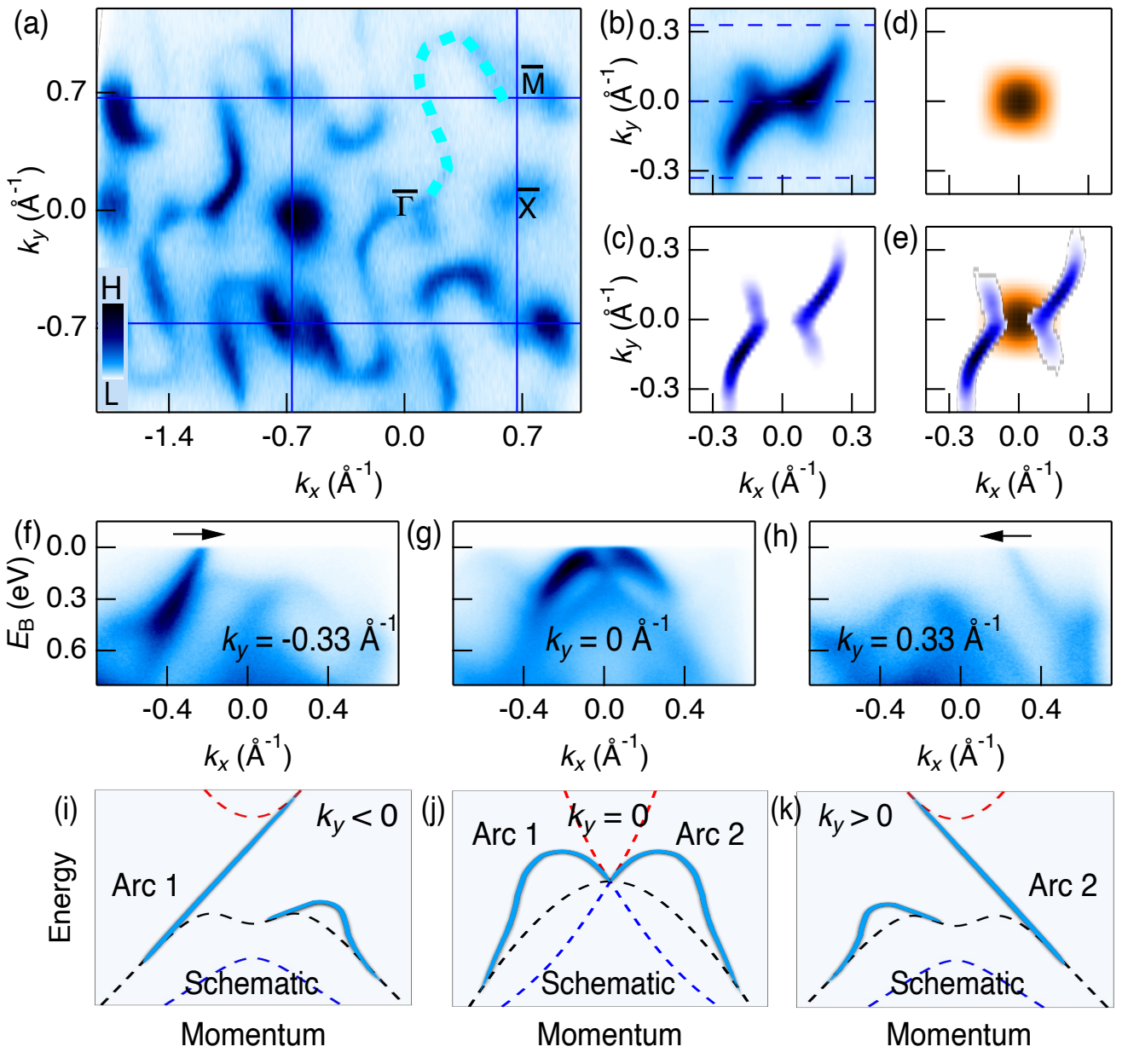

FIG. 3. Fermi arc switching in momentum space. (a) Surface sensitive UVARPES measured Fermi surface from (001) surface, utiliizing 85 eV incident photons. Long Fermi arc surface states connect pockets at $\bar{\Gamma}$ and $\bar{\mathrm{M}}$ as indicated by the dashed cyan line. (b) UVARPES measured Fermi surface at the $\bar{\Gamma}$ point with 40 eV incident photons. (c) Second derivative of spectrum in Fig. 3(b). (d) SXARPES measured Fermi surface at the Γ point. (e) Overlay of Fig. 3(c) on Fig. 3(d) showing the Fermi arcs connecting to the bulk band. (f)-(h) $E_\mathrm{B}$ vs $k$ UVARPES spectra showing chiral Fermi arc surface states along paths defined by the dotted lines in Fig. 3(b). Arrows indicate chiral modes at the Fermi level, which switch at the three-fold chiral fermion at the $\bar{\Gamma}$ point. (i)-(k), Schematics of chiral Fermi arcs (solid blue lines) and bulk bands (dotted lines) depicting the Fermi arc switching at the $\bar{\Gamma}$ point.

To gain additional insight into the three-fold chiral fermion, we examine constant $E_\mathrm{B}$ contours. At $E_\mathrm{B} = 400$ meV, below the degeneracy point, two bands are observed [Figs. 2(e) and 2(f)]. Band 1 is circular, whereas band 2 has an apparent distortion, which explains the variation between the band separations along the Γ-X and Γ-M directions. Above the degeneracy point at the Fermi level, only one band is observed; it is circular. Therefore, bands 1 and 3 form a nearly isotropic cone, whereas band 2 has a parabolic-like dispersion that is distorted by the cubic crystal symmetry. These results allow us to conclude we have observed all three branches of a three-fold chiral fermion at the Γ point.

Having characterized the three-fold bulk crossing, we now present experimental evidence of its nontrivial higher-fold topology. Using surface-sensitive ultraviolet angle-resolve photoemission spectroscoopy (UVARPES), we directly observe topological surface states in two gapped regions of the bulk electronic structure. By utilizing 85 eV incident photons, the measured (001) Fermi surface covers multiple surface Brillouin zones [Fig. 3(a)]. Long states are observed stretching across each Brillouin zone from the $\bar{\Gamma}$ point to the $\bar{\mathrm{M}}$ point. Further decreasing the photon energy to 40 eV to improve energy and momentum resolution, these states disperse along the same trajectory. Focusing near the $\bar{\Gamma}$ point, we observe the surface states have a high photoemission intensity [Fig. 3(b)]. By taking the second derivative of the spectrum, we see the states are discontinuous at the $\bar{\Gamma}$ point [Fig. 3(c)]. To explain this, we turn to the bulk spectrum [Fig. 3(d)]. Overlaying the surface onto the bulk dispersion, it is evident that the surface states emanate from the bulk bands as stipulated by the bulk-boundary correspondence for Fermi arcs [Fig. 3(e)]. Further understanding of these states can be gained by analyzing their $E_{\mathrm{B}}$ vs $k$ dispersion near the $\bar{\Gamma}$ point. At negative $k_y$, one isolated right-moving chiral mode is observed, which corresponds to a Chern number of $C(k_y < 0) = 1$ along the bulk plane that projects onto the line $k_y = -0.33$ Å$^{-1}$ [Fig. 3(f)]. At $k_y = 0$ Å$^{-1}$, the chiral state and its time-reversal partner connect directly to the projection of the bulk degeneracy point, showing the surface to bulk connectivity [Fig. 3(g)]. Moving to positive $k_y$, one isolated left-moving chiral mode is observed, indicating $C(k_y > 0) = -1$ on the corresponding bulk plane [Fig. 3(h)]. Together, these three dispersions provide strong evidence that the surface states are topological Fermi arcs, connecting bands 2 and 3 of the bulk three-fold chiral fermion. Moreover, these measurements reveal the detailed nature of Fermi arc switching at a higher-fold fermion. We clearly see a surface state transition from topological (connecting bands 2 and 3) to trivial (starting and ending at band 2) as the $E_{\mathrm{B}}$ vs $k$ dispersion is scanned across the degeneracy point, as illustrated in the schematics [Figs. 3(i), 3(j), and 3(k)]. To extract the chiral charge for gap 2, we note that the difference between the Chern numbers along two lines is equal to the chiral charge enclosed by those lines [51]. In this case, we obtain that the chiral charge in gap 2 is $C_2 = C(k_y < 0) - C(k_y > 0) = 1 - (-1) = 2$.

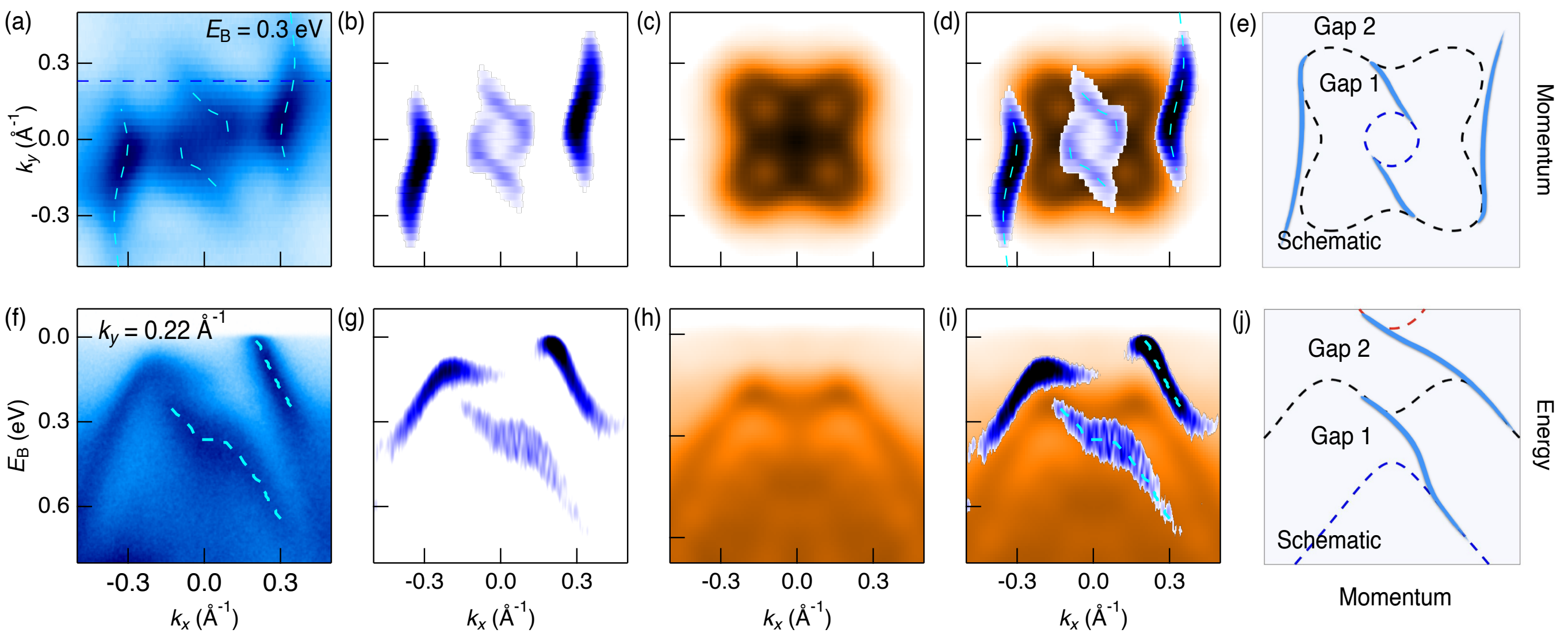

FIG. 4. Multi-gap chiral charge and a Fermi arc ladder. (a),(b) UVARPES measured constant $E_{\mathrm{B}}$ contour at $E_{\mathrm{B}} = 300$ meV [Fig. 4(a)] and second-derivative [Fig. 4(b)] at the $\bar{\Gamma}$ point. Both inner and outer states are indicated by cyan dashed lines in Fig. 4(a). (c) SXARPES constant $E_{\mathrm{B}}$ contour at $E_{\mathrm{B}} = 300$ meV at the $\Gamma$ point. (d) Overlay of Fig. 4(b) onto Fig. 4(c); the inner surface states lie within the bulk projection of band 2. (e) Schematic constant $E_{\mathrm{B}}$ contour showing chiral states (thick blue lines) in gap 1 and gap 2 of the bulk three-fold chiral fermion (dotted lines). (f),(g) UVARPES measured $E_{\mathrm{B}}$ vs $k$ cut at $k_y = 0.22$ Å$^{-1}$ [Fig. 4(f)] and second-derivative [Fig. 4(g)], along the path indicated by the dark blue dashed line in Fig. 4(a). Two disconnected chiral modes are observed (dashed cyan lines). (h) SXARPES measured spectrum along the same path as Fig. 4(f) at $k_y = 0.22$ Å$^{-1}$. (i) Overlay of Fig. 4(g) onto Fig. 4(h), showing topological surface states propagating in gap 1 and 2 respectively. (j) Schematic $E_{\mathrm{B}}$ vs $k$ plot showing the Fermi arc ladder (thick blue lines).

To determine the topology of gap 1, we examine the band structure below the Fermi level. A constant $E_{\mathrm{B}}$ contour measured at $E_{\mathrm{B}} = 300$ meV shows signatures of multiple chiral modes indicated by dashed cyan lines [Fig. 4(a)]. While the Fermi arcs in gap 2 are still present near the edge of Fig. 4(a), there is another set of states residing closer to $\bar{\Gamma}$. To better resolve these states, we take the second derivative of the constant $E_{\mathrm{B}}$ spectrum [Fig. 4(b)], and observe that the inner states are disconnected and distinct from the outer Fermi arcs. To know which gap to associate these surface state with, we contrast the UVARPES and the bulk SXARPES spectra [Fig. 4(c)]. By comparing the surface states and bulk constant $E_{\mathrm{B}}$ contours [Fig. 4(d)], we see that the inner surface states propagate from band 1 to band 2, consistent with these states being topological Fermi arcs in gap 1 [Fig. 4(e)]. To confirm this, we measure $E_{\mathrm{B}}$ vs $k$ UVARPES spectra along the path indicated by the dashed, dark blue line in Fig. 4(a). We identify two distinct chiral modes [Fig. 4(f)]. Overlaying the second derivative of the surface state spectrum [Fig. 4(g)] onto the bulk spectrum [Fig. 4(g)], we see that one chiral mode lies entirely within gap 1, while the other disperses in gap 2, forming a ladder structure [Fig. 4(i)]. The state in gap 2 is the Fermi arc already discussed in Fig. 3. The chiral state in gap 1 is a second Fermi arc, connecting bands 1 and 2. With an identical procedure to that carried out for gap 2, the chiral charge of gap 1 can be experimentally assigned to be $C_1 = 2$. Together our measurements in gap 1 and 2 visualize a Fermi arc ladder, which demonstrates the multi-gap bulk boundary correspondence for a higher-fold chiral fermion with multi-gap chiral charge $\mathbf{C} = (2, 2)$ [Fig. 4(j)].

Through the multi-gap bulk-boundary correspondence, we have discovered higher-fold topology in a three-fold chiral fermion, with multi-gap chiral charge $\mathbf{C} = (2, 2)$. This work further motivates transport and optical research on higher-fold topological materials. Indeed, our material-realistic theoretical simulations presented in the Supplemental Materials reveal a giant enhancement to optical sum frequency response, unique to crystals with multiple topological inter-band gaps [52–57]. This stems from optical resonances across multiple topological gaps, which is only possible in the multi-gap regime. Furthermore, by considering our chemical engineering in the simulations we elucidate the possibility of unprecedented photocurrent response of our samples. Our technique of freely doping the Fermi level and preserving the topological bands provides a concrete pathway to realize a tunable quantized circular photogalvanic effect, which has been elusive in pristine RhSi and CoSi samples [58–60]. Specifically, we predict quantized photocurrent behavior not only in our $Rh_{1-x}Ni_xSi$ samples, but also in $Rh_{1-x}Fe_xSi$. These results underscore the importance of the present study, while motivating further transport and optical research on higher-fold topological materials. For example, we expect exotic behavior when considering inter-arc and arc-to-bulk/bulk-to-arc transitions in RhSi, inducing unconventional surface photocurrents with applications to thin film devices [21]. Furthermore, the multi-gap topology presented here is not restricted to electronic crystals, and can be realized in mechanical [61], phononic [62], photonic [63], cold atom [64], qubit [65], and even atmospheric systems [66], opening up opportunities for future research.

The authors acknowledge Songtian S. Zhang and Nana Shumiya for useful conversations during manuscript preparations. The authors also acknowledge Takayuki Muro for beamline support at SPring-8 BL25SU. Work at Princeton University and Princeton-led synchrotron based ARPES measurements were supported by the United States Department of Energy (US DOE) under the Basic Energy Sciences program (grant number DOE/BES DE-FG-02-05ER46200). This research used resources of the Advanced Light Source, which is a DOE Office of Science User Facility under contract No. DE-AC02-05CH11231. Synchrotron radiation experiments were performed at the BL25SU of SPring-8 with the approval of the Japan Synchrotron Radiation Research Institute (JASRI) (Proposal No. 2018A1684 & 2019A1696). We acknowledge the Paul Scherrer Institut, Villigen, Switzerland for provision of synchrotron radiation beamtime at the ADRESS beamline of the Swiss Light Source. We acknowledge SOLEIL for provision of synchrotron radiation facilities at the CASSIOPÉE beamline. Additional ARPES measurements were performed at the RGBL-2 end station at the U125/2 undulator beamline of BESSY II. T. A. C. was supported by the National Science Foundation Graduate Research Fellowship Program under Grant No. DGE-1656466. J. S.-B. gratefully acknowledges financial support from the Impuls-und Vernetzungsfonds der Helmholtz-Gemeinschaft under grant No. HRSF-0067 (Helmholtz-Russia Joint Research Group). K. M. and C. F. thank the financial support of the European Research Council (ERC) with Advanced Grant No. (742068) "TOP-MAT". J. A. K. acknowledges support from the Swiss National Science Foundation (SNF-Grant No. 200021_165910). G.C. is supported by the National Research Foundation, Singapore under its NRF Fellowship Award (NRF-NRFF13-2021-0010) and the Nanyang Assistant Professorship grant from Nanyang Technological University. T.A.C., I.B., and G.C. contributed equally to this work.

# Supplemental Material for
# Visualizing higher-fold topology in chiral crystals

Tyler A Cochran[*,†,1] Ilya Belopolski[*,1] Kaustuv Manna,[2,3] Mohammad Yahyavi,[4] Yiyuan Liu,[5] Daniel S. Sanchez,[1] Zi-Jia Cheng,[1] Xian P. Yang,[1] Daniel Multer,[1] Jia-Xin Yin,[1] Horst Borrmann,[2] Alla Chikina,[6] Jonas A. Krieger,[6,7] Jaime Sánchez-Barriga,[8] Patrick Le Fèvre,[9] François Bertran,[9] Vladimir N. Strocov,[6] Jonathan D. Denlinger,[10] Tay-Rong Chang,[4] Shuang Jia,[5] Claudia Felser,[2] Hsin Lin,[11] Guoqing Chang[†,12] and M. Zahid Hasan[†1,13,14]

[1]*Laboratory for Topological Quantum Matter and Advanced Spectroscopy (B7), Department of Physics, Princeton University, Princeton, New Jersey 08544, USA*

[2]*Max Planck Institute for Chemical Physics of Solids, Dresden, Germany*

[3]*Department of Physics, Indian Institute of Technology Delhi, Hauz Khas, New Delhi 110016, India*

[4]*Department of Physics, National Cheng Kung University, Tainan 70101, Taiwan*

[5]*International Center for Quantum Materials, School of Physics, Peking University, Beijing, China*

[6]*Swiss Light Source, Paul Scherrer Institute, Villigen, Switzerland*

[7]*Laboratory for Muon Spin Spectroscopy, Paul Scherrer Institute, Villigen, Switzerland*

[8]*Helmholtz-Zentrum Berlin für Materialien und Energie, Elektronenspeicherring BESSY II, Albert-Einstein Strasse 15, 12489 Berlin, Germany*

[9]*Synchrotron SOLEIL, L'Orme des Merisiers, Saint-Aubin-BP 48, 91192 Gif-sur-Yvette, France*

[10]*Advanced Light Source, Lawrence Berkeley National Laboratory, Berkeley, California 94720, USA*

[11]*Institute of Physics, Academia Sinica, Taipei 11529, Taiwan*

[12]*Division of Physics and Applied Physics, School of Physical and Mathematical Sciences, Nanyang Technological University, 21 Nanyang Link, 637371, Singapore*

[13]*Princeton Institute for Science and Technology of Materials,*

*Princeton University, Princeton, New Jersey 08544, USA*
[14]*Lawrence Berkeley National Laboratory, Berkelely, California 94720, USA*

## CONTENTS

# I. SUPPLEMENTAL TEXT

## A. Previous research on topological chiral crystals

While work has been performed on the electronic topological properties of chiral crystals in space group $P2_13$ (#198), conclusive evidence of higher-fold topology is still lacking. *Ab initio* studies have identified a number of compounds that should host higher-fold chiral fermions [18, 33, 34]. Also, experimental efforts have been carried out to investigate large Chern numbers [24–27]. However, none of these studies have demonstrated the two key properties of higher-fold topology: (1) Resolving all the bulk bands that become degenerate at a higher-fold chiral fermion, and (2) showing nontrivial Chern number in each inter-band gap.

The chemical engineering of $Rh_{1-x}Ni_xSi$ in this work is essential to fulfilling these criteria. In undoped RhSi, the three-fold degeneracy at the $\Gamma$ is above the Fermi level [Fig. S1(a) and S1(b)]. This means the uppermost branch of the higher-fold chiral fermion is entirely above the Fermi level, and therefore completely undetectable by conventional photoemission experiments. Also in RhSi, the branches of the four-fold degeneracy at the R point are not split sufficiently to resolve all branches, leaving the crossing degeneracy and higher-fold topology unaddressed [Fig. S1(f) and S1(g)]. Additionally, state-of-the-art UVARPES measurements have been unable to demonstrate Fermi arc states in more than one inter-band gap. Thus the existence of higher-fold chiral fermions from previous ARPES measurements is impossible, as criterion (1) and (2) cannot be met. Similarly, in no previous work on any material have the listed criteria been demonstrated to prove higher-fold topology [Fig. S1].

### B. Single crystal characterization

The single crystallinity was first checked with a white beam backscattering Laue X-ray setup at room temperature. A picture of the grown single crystal and the refined Laue pattern is shown in Fig. S2(a) and S2(b). In order to verify the presence of Ni, Rh, and Si in the grown single crystal, we first performed the compositional analysis with energy-dispersive X-ray (EDX) spectroscopy and a representative plot is shown in Fig. S2(b), clearly revealing the presence of all three elements in the compound. Then a detailed chemical analysis was carried out to precisely determine the chemical composition of the grown single crystal and is estimated as $Rh_{0.955}Ni_{0.045}Si$.

A small piece was broken from the large crystal and selected fragments were mounted on Kapton loops and tested for structure refinement. Data were collected on a Rigaku AFC7 four-circle diffractometer with a Saturn 724+ CCD-detector applying graphite-monochromatized Mo-K$\alpha$ radiation. The pertinent details are given in Table S1. Due to strong correlations, a free refinement of occupancies at the mixed Rh/Ni site did not give very reasonable results as indicated in particular by comparison of displacement parameters at Si versus Rh/Ni atom site. Since in binary RhSi counterintuitively the displacement parameter of Si is slightly smaller than at the other site, occupancies for the metals were optimized using this condition as a main guide. Along this line, an optimized model indicates 7% Ni at the metal site which is in quite good agreement with analytical results, even more when considering large difference in scattering power of Ni vs. Rh. The absolute structure of the crystal clearly has to be assigned to B-form [42] and the Flack parameter refined to -0.01(11). Evidently, the refined Flack's parameter confirms single chirality domain formation in the grown single crystal. As expected the absolute structure refinement is not particularly sensitive to variation of occupancies at the metal site.

| Compound | $Rh_{1-x}Ni_xSi$ |
|---|---|
| F.W. (g/mol); | 127.91 |
| Space group;Z | $P2_13$ (No.198);4 |
| $a$ (Å) | 4.6766(6) |
| $V$ ($Å^3$) | 102.28(4) |
| Absorption Correction | Multi-scan |
| Extinction Coefficient | 0.22(1) |
| $\theta$ range (deg) | 4.4-36.8 |
| No. independent reflections | 165 |
| No. parameters | 8 |
| $R_1$; $wR_2$ (allI) | 0.0212; 0.0464 |
| Goodness of fit | 1.122 |
| Diffraction peak and hole ($e^-/Å^3$) | 1.188;-1.476 |
| Refined formula | $Rh_{0.93}Ni_{0.07}Si$ |

Table S1. Single crystalline x-ray diffraction of $Rh_{1-x}Ni_xSi$.

The magnetization measurement was performed using a Quantum Design vibrating sample magnetometer (MPMS) with magnetic field up to 7 T. The diamagnetic behavior of the $Rh_{1-x}Ni_xSi$ single crystal is evident from the field dependent magnetization data at various temperatures 5, 20 and 50 K [Fig. S2(c)]. Similar diamagnetic behavior is also observed for the Weyl fermions in TaAs [43] and the Dirac fermions in Bi [44, 45]. The field-independent negative dc-susceptibility throughout the measured temperature range indicates the absence of any long range magnetic ordering and confirms the intrinsic diamagnetism in our $Rh_{1-x}Ni_xSi$ single crystals [Fig. S2(d)].

### C. Band structure tuning in $Rh_{1-x}Ni_xSi$

The substitution of Ni into the RhSi lattice is critical to the observation of higher-fold topology discussed in this work. In pristine RhSi samples, no band degeneracy is observed below the Fermi level [Fig. S3(a)]. Also, the top-most band is not observed anywhere in ARPES intensity near the Γ point. Therefore, the existence of a chiral fermion can only be inferred by theoretical predictions [18, 33, 34]. By tuning the band structure with the Ni substitution, we are able to observe the degeneracy point between all three bands at the Γ point [Fig. S3(b)]. This is the first time all bands of a higher-fold chiral fermion have been resolved. Our ARPES data suggests that the primary change in the electronic structure is equivalent to an effective shift in the chemical potential of 300 meV [Fig. S3(c)].

### D. Multi-gap topology drives giant sum frequency response in RhSi

Apparent in our discovery of higher-fold topology is the central role of multiple topological inter-band gaps. Considering this, we use state-of-the-art material specific simulation to explore the implications that multi-gap topology may have on electromagnetic response functions, beyond conventional two-fold Weyl fermions (Supplemental Material, Sec. II C). Of all the possible response functions that may be sensitive to multiple topological gaps, we choose to explore the optical conductivity. This is a fundamental quantity to describe the propagation and absorption of light in materials, closely related to the dielectric function. We investigate the second-order nonlinear optical conductivity, $\sigma^{\alpha\beta;\gamma}$, which relates the current induced in a material by two distinct oscillating electric fields, often in the form of laser light: $J_\gamma(\omega) = \sigma^{\alpha\beta;\gamma}(\omega, \omega_1, \omega_2) E_\alpha(\omega_1) E_\beta(\omega_2)$ with $\omega = \omega_1 + \omega_2$. Focusing on the case where $\omega_1, \omega_2 > 0$, $\sigma^{\alpha\beta;\gamma}$ is then related to sum-frequency generation: when two lasers of frequency $\omega_1$ and $\omega_2$ are applied to the sample, an output beam of frequency $\omega$ is produced. One contribution to such a process only involves two bands, analogous to the case of a Weyl fermion [Fig. S4(a) black arrows]. However, there is also a contribution that involves three bands [Fig. S4(a) cyan arrow]. Our photoemission studies on $Rh_{1-x}Ni_xSi$ reveal the possible optical transitions that could give rise to this three-band mechanism in RhSi [Fig. S4(b) and S4(c)]. Unexpectedly, by calculating $\sigma^{\alpha\beta;\gamma}$, we find that the three-band mechanism dominates the second-order optical conductivity of the three-fold chiral fermion in RhSi, which is one order larger than the conventional two-band contributions [Fig. S4(d)]. Such results indicate that giant sum-frequency generation response in topological semimetals is further greatly enhanced in topological multi-gap systems [53, 54].

### E. Quantized CPGE in $Rh_{1-x}Ni_xSi$ and $Rh_{1-x}Fe_xSi$

Having visualized higher-fold topology in the electronic structure of $Rh_{1-x}Ni_xSi$, we generally expect unconventional electromagnetic response related to the nontrivial topology. Of all the possible response functions that could exemplify the applicability of higher-fold topology (beyond sum-frequency generation), we choose to explore the circular photogalvanic effect (CPGE), which previous theoretical studies have indicated may give rise to a quantized injection current proportional to the chiral charge of chiral fermions [17, 18, 33].

Previous THz emission spectroscopy studies have probed ultrafast photocurrents in RhSi and CoSi [19, 59, 60]. While results initially indicated a potential CPGE plateau in RhSi, the quantized value of the plateau was not realized [19]. Therefore, the connection of these photocurrents to the chiral charge of the three-fold fermion could not be established. Further experiments on RhSi and CoSi also did not uncover a quantized plateau in the CPGE signal, indicating the possibility of a Fermi level misalignment in the samples [59, 60].

Having extracted the Fermi level of the samples with our photoemission measurements, we are able to calculate the CPGE response in the native RhSi crystals and compare to the previous results [Figs. S5 and S6(a)]. In our simulation, the signal exhibits a non-quantized negative peak between 500-600 meV. We believe this feature corresponds to the positive peak initially measured by Rees *et al.*, since THz emission spectroscopy is not sensitive to the absolute sign of photocurrents. Our simulation further predicts that no quantized plateau should exist in undoped RhSi in the energy range probed by experiments.

Prompted by our chemical gating and photoemission studies of RhSi, we also simulated the CPGE spectrum for electron-doped $Rh_{1-x}Ni_xSi$ and holed-doped $Rh_{1-x}Fe_xSi$ [Supplemental Materials, Sec. II C]. In $Rh_{1-x}Ni_xSi$ samples, the $+2\ e^2/h^3$ quantized plateau is present in a large energy range between 260-690 meV [Fig. S6(b)]. Furthermore, in $Rh_{1-x}Fe_xSi$ samples, a $+2\ e^2/h^3$ quantized plateau is observed at energies between 120-200 meV, while another plateau with value $-2\ e^2/h^3$ spans between 480-820 meV [Fig. S6(c)]. Having determined the detailed electronic structure of the bulk bands, we are in a position to elucidate the correspondence between the bulk band structure and the CPGE response [Fig. S6(d)]. Our analysis reveals the plateau in $Rh_{1-x}Ni_xSi$ samples is due to optical transitions across gap 2 near the $\Gamma$ point, showing that $C_2 = +2$. Furthermore, the plateaus in $Rh_{1-x}Fe_xSi$ are due to transitions across gap 1 and gap 2 at the $\Gamma$ and R points, respectively, showing $C_1 = +2$ at the $\Gamma$ point. Therefore, the bulk-photogalvanic correspondence between our SXARPES and numerical CPGE experiments confirms our assignment of the multi-gap chiral charge: $\mathbf{C} = (2, 2)$.

These results provide a clear path to realize the sought-after quantized CPGE in experiments. The $Rh_{1-x}Ni_xSi$ samples we have engineered exhibit a large and robust quantized CPGE plateau in an energy range accessible to optical experiments. Also, by further tuning the samples with Fe substitution, a regime can be achieved with a $C = +2$ plateau from gap 1 at the $\Gamma$ point and a $C = -2$ plateau from gap 2 at the R point. Such a highly tunable, quantized photocurrent response is possible in real samples, as our Fe substitution into the RhSi family shows [Supplementary Materials, Sec I F].

## II. MATERIALS AND METHODS

### A. Single crystal growth

A single crystal of the $Rh_{1-x}Ni_xSi$ was grown from the melt using the vertical Bridgman crystal growth technique. First, polycrystalline ingot of the Ni-substituted composition, $Rh_{0.95}Ni_{0.05}Si$ was prepared by pre-melting the highly pure stoichiometric amount of respective metals under argon atmosphere using an arc melt technique. Then the crushed powder was filled in a custom-designed sharp-edged alumina tube, which was again sealed inside a tantalum tube with argon atmosphere. Here we induced a slight excess of Si in the composition to ensure a flux growth inside the Bridgman ampule. First, the sample was heated to 1550° C with a rate of 200º C/h and held there for 10 h. Then the ampule was slowly pulled to the cold zone, down to 1100° C with a rate of 0.8 mm/h. The temperature was controlled by attaching a thermocouple at the bottom of the Bridgman ampule. A single crystal with average dimension of 9 mm length and 6 mm diameter was obtained.

### B. Angle-resolved photoemission spectroscopy

Directly before photoemission measurements, (001) polished single crystals were *in situ* sputtered with 500-1000 eV Ar ions for 5-30 minutes and annealed to 550-585° C (assuming a sample emissivity of 0.45) for 30-45 minutes in ultra-high vacuum (UHV), cyclically.

1. ***SXARPES studies*** on $Rh_{1-x}Ni_xSi$ were carried out with incident light in the range 400-750 eV at beamline BL25SU of SPring-8 using a Scienta DA30 deflection-based hemispherical analyzer. Sample temperature was maintained between 21-25 K during all ARPES measurements and vacuum conditions were UHV. The increased photoelectron mean free path in the soft X-ray energy range results, by the Heisenberg uncertainty principle, in high intrinsic $k_z$ resolution of the ARPES experiment important for three-dimensional materials [47].

2. ***SXARPES studies*** on RhSi were carried out with incident light in the range 400-750 eV at the ADRESS beamline of the Swiss Light Source using a Specs PHOIBIOS-150 analyzer. Sample temperature was maintained below 20 K during all ARPES measurements and vacuum conditions were UHV [48–50].

3. ***UVARPES studies*** were carried out at beamline 4.0.3 of the Advanced Light Source using a Scienta R8000 hemispherical analyzer and incident light in the range 40-85 eV. Sample temperature was maintained between 17-20 K during all measurements and vacuum conditions were UHV.

### C. *Ab initio* calculation

First-principles calculations were performed within the density functional theory (DFT) framework using the projector augmented wave method as implemented in the VASP package [55, 56]. A Γ-centered k-point 17x17x17 mesh was used.

The circular photogalvanic effect (CPGE) injection current was calculated using the equation [17]

$$\frac{dj_i}{dt} = \beta_{ij}(\omega)\left[\mathbf{E}(\omega)\times\mathbf{E}^*(\omega)\right]^j , \tag{1}$$

where $\mathbf{E}(\omega) = \mathbf{E}^*(-\omega)$ is the electric field. The CPGE injection current tensor is defined as [17]

$$\beta_{ij}(\omega) = \frac{\pi e^3}{\hbar^2}\epsilon_{j\beta\gamma}\int_{\mathbf{k}}\sum_{n,m} f_{nm}\Delta^i_{mn} r^\beta_{nm} r^\gamma_{mn}\delta(\varepsilon_{mn}-\hbar\omega) \tag{2}$$

where $\int_{\mathbf{k}} = \int d^dk/(2\pi)^d$, $\varepsilon_{mn} = \varepsilon_m - \varepsilon_n$ is the energy difference between bands $n$ and $m$, $f_n$ is the Fermi-Dirac distribution of the band $n$, $f_{nm} = f_n - f_m$, $\Delta^i_{mn} = \partial_{k_i}\varepsilon_{nm}/\hbar$, and $r^\beta_{nm} = i\langle u^n_{\vec{k}}|\partial_{k_\beta}|u^m_{\vec{k}}\rangle$ is the off-diagonal Berry connection or interband transition matrix element.

To calculate the sum frequency response, we use general form of the nonlinear optical conductivity response of higher-fold chiral fermions, where two photons are absorbed in the transitions (Fig. S1(a) black arrows). For the clean system, the complete second-order conductivity takes the form [57]

$$\begin{aligned}\sigma^{\alpha\beta;\gamma}(\omega,\omega_1,\omega_2) = \frac{-e^3}{\hbar^2\omega_1\omega_2}\sum_{m,n,p}\int_{\boldsymbol{k}} & f_m u^{\alpha\beta\gamma}_{mm} \\ & + f_{mn}\frac{v^{\alpha}_{mn}w^{\gamma\beta}_{nm}}{\omega_1-\varepsilon_{mn}} + f_{mn}\frac{v^{\beta}_{mn}w^{\gamma\alpha}_{nm}}{\omega_2-\varepsilon_{mn}} + f_{mn}\frac{w^{\alpha\beta}_{mn}v^{\gamma}_{nm}}{\omega-\varepsilon_{mn}} \\ & + \left(f_{mn}\frac{v^{\alpha}_{mn}v^{\beta}_{np}v^{\gamma}_{pm}}{(\omega_1-\varepsilon_{nm})(\omega-\varepsilon_{pm})} + f_{pn}\frac{v^{\alpha}_{mn}v^{\beta}_{np}v^{\gamma}_{pm}}{(\omega_2-\varepsilon_{pn})(\omega-\varepsilon_{pm})} + (\alpha,\omega_1 \leftrightarrow \beta,\omega_2)\right),\end{aligned} \tag{3}$$

where

$$\begin{aligned} v^{\alpha}_{mn} &= \langle m|\partial^{\alpha}\hat{H}(\boldsymbol{k})|n\rangle \\ w^{\alpha\beta}_{mn} &= \langle m|\partial^{\alpha}\partial^{\beta}\hat{H}(\boldsymbol{k})|n\rangle \\ u^{\alpha\beta\gamma}_{mn} &= \langle m|\partial^{\alpha}\partial^{\beta}\partial^{\gamma}\hat{H}(\boldsymbol{k})|n\rangle \end{aligned} \tag{4}$$

where the frequency of the emitted photon is $\omega = \omega_1 + \omega_2$, $(\alpha, \beta, \gamma) = (x, y, z)$ and where the sum over band indices is only performed over the indices appearing in each term.

To understand the nonlinear conductivity response in higher-fold chiral fermion systems, here, we calculate the nonlinear conductivity $\sigma^{yz;x}$ as an example. We fix the frequency of one photon $\hbar\omega_1 = 200$ meV and scan the frequency of the second $\omega_2$. In particular, we separate the contributions of the two- and three-band processes for the nonlinear response of higher-fold chiral fermions. We can see that the second-order conductivity for the two-band process (fig. S1D black and grey traces) is generally at least 1 order of magnitude smaller than the three-band process (fig. S1D blue trace). This result suggests that three-fold chiral fermions will intrinsically possess much larger nonlinear responses compared to conventional two-fold Weyl fermions.

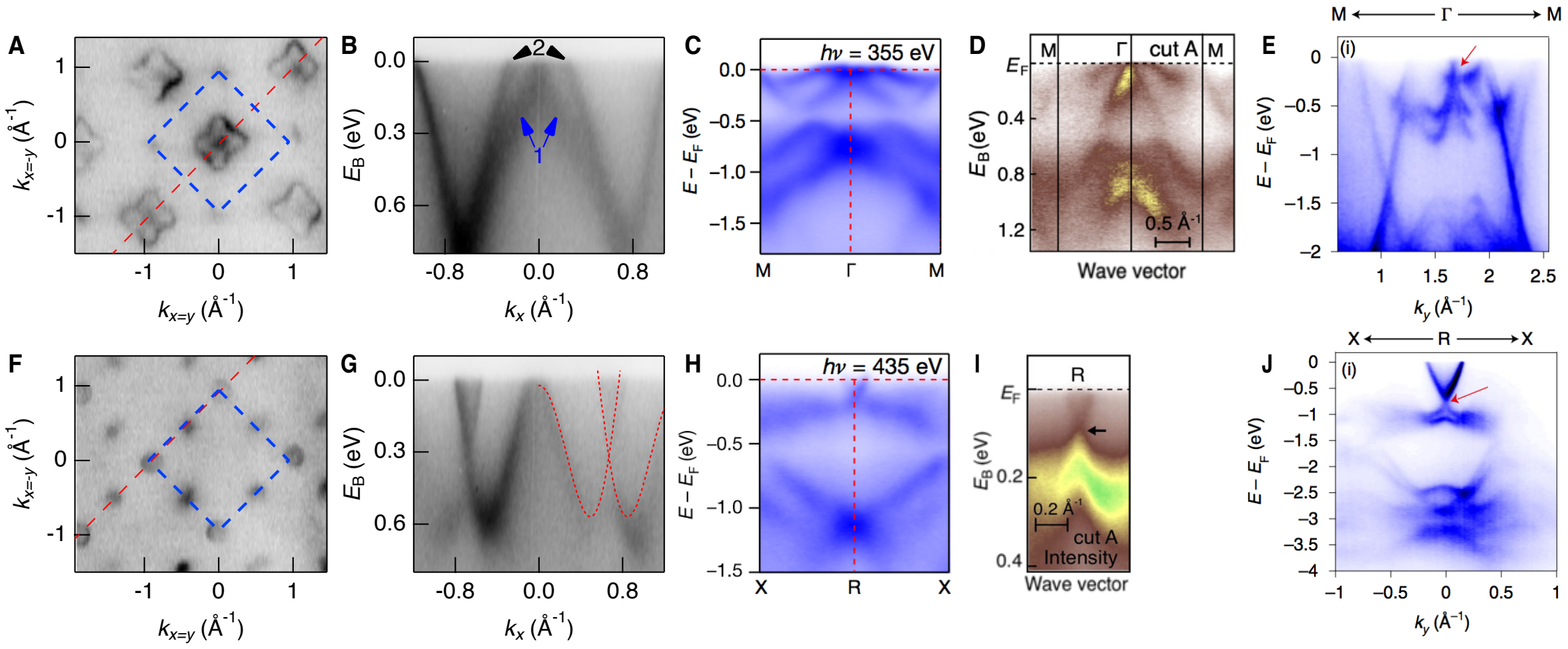

FIG. S1. SXARPES measurements on RhSi and previously published data inconclusive for higher-fold chiral fermions. (a) RhSi Fermi surface measured at the $k_z = 0$ plane with a photon energy of 540eV. The first Brillouin zone boundary is outlined by the dashed blue line. (b) $E_B$ vs $k$ cut along the dashed red line in (a), passing through the $\Gamma$ point of RhSi. Two branches are observed, but no third branch or higher-fold degeneracy. (c) $E_B$ vs $k$ cut passing through the $\Gamma$ point of CoSi, adapted from ref [25]. Two branches are observed, but no third branch or higher-fold degeneracy. (d) $E_B$ vs $k$ cut passing through the $\Gamma$ point of CoSi, adapted from ref [27]. Two branches are observed, but no third branch or higher-fold degeneracy. (e) $E_B$ vs $k$ cut passing through the $\Gamma$ point of AlPt, adapted from ref [26]. All bands of the 4-fold degeneracy cannot be resolved. (f) Fermi surface of RhSi measured at the $k_z = \pi$ plane with a photon energy of 472 eV. The first Brillouin zone boundary is outlined by the dashed blue line. (g) $E_B$ vs $k$ cut along the dashed red line in **c**, passing through the R point of RhSi. The data shows a linear crossing, but only two branches of the four-fold fermion can be resolved within experimental resolution. The dotted red line shows the shape of the band dispersion measured by ARPES. (h) $E_B$ vs $k$ cut passing through the R point of CoSi, adapted form ref [25]. The four branches of the higher-fold fermion are not resolved. (i) $E_B$ vs $k$ cut passing through the R point of CoSi, adapted from ref [27]. The four branches of the higher-fold fermion are not resolved. (j) $E_B$ vs $k$ cut passing through the R point of AlPt, adapted from ref [26]. The six branches of the higher-fold fermion are not resolved.

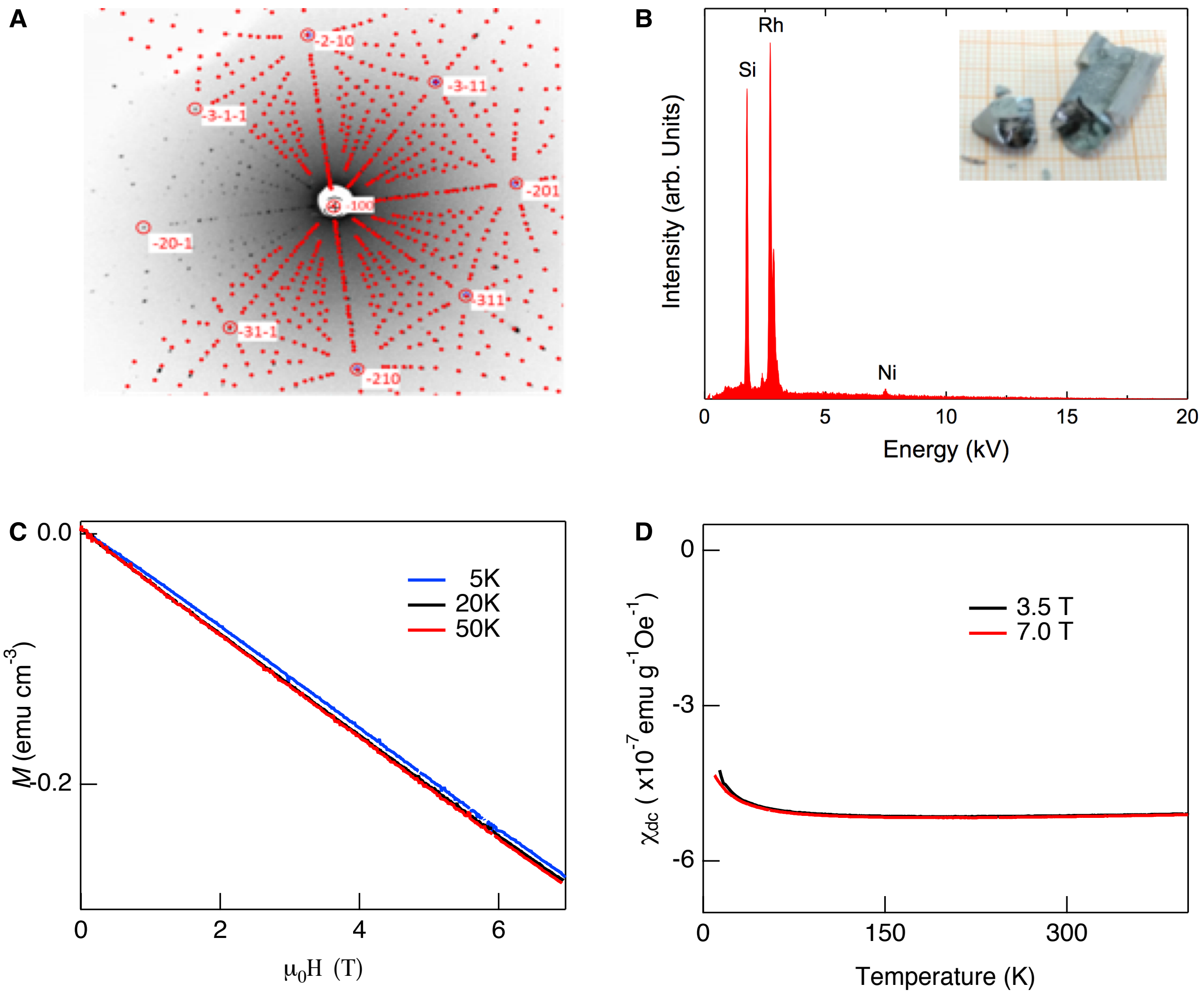

FIG. S2. Physical characterization of $Rh_{1-x}Ni_xSi$ single crystals. (a) Laue diffraction pattern of a [100] oriented crystal superposed with a theoretically simulated pattern confirming high crystal quality. (b) EDX pattern of the $Rh_{1-x}Ni_xSi$ single crystals. Picture of the grown $Rh_{1-x}Ni_xSi$ single crystal (inset). (c) Magnetization measured at 5, 20, and 50 K. (d) Magnetic susceptibility measured with 3.5 and 7 T fields, respectively.

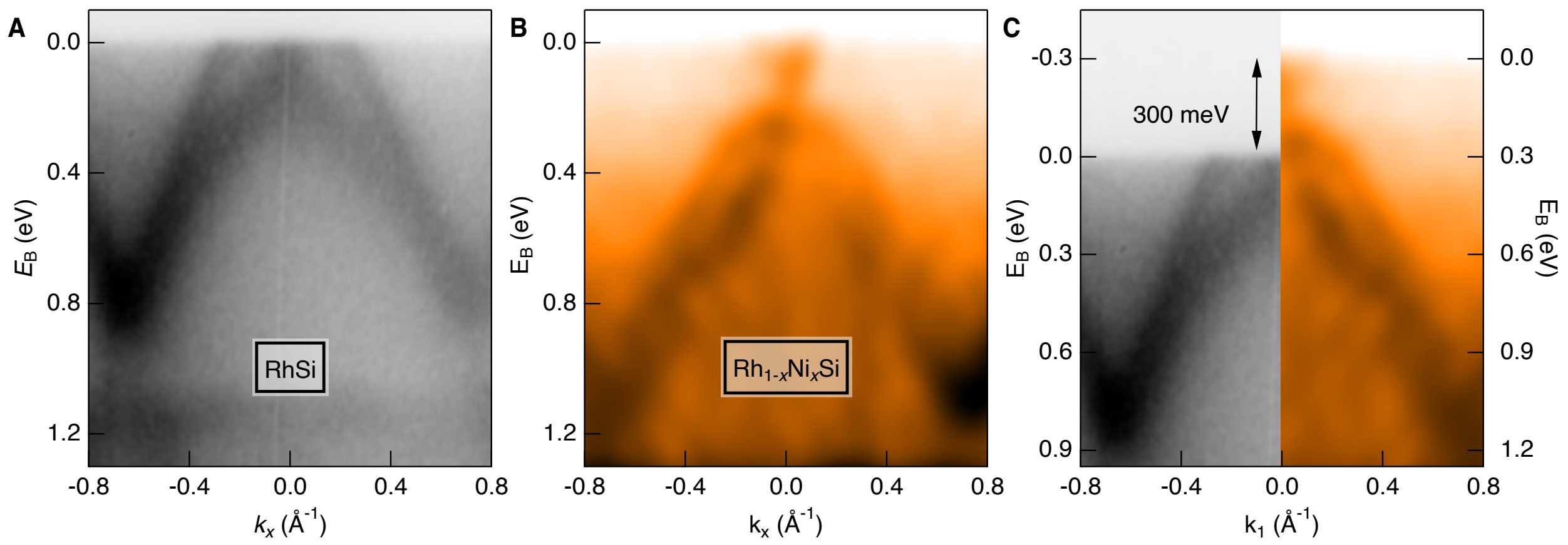

FIG. S3. Comparison of ARPES spectra of RhSi and $Rh_{1-x}Ni_xSi$. (a) ARPES spectrum of RhSi taken using 540 eV incident photons along the Γ-X direction. (b) ARPES spectrum of $Rh_{1-x}Ni_xSi$ taken using 550 eV incident photons along the Γ-X direction. (c) Comparison between RhSi and $Rh_{1-x}Ni_xSi$ spectra, showing an effective shift of the chemical potential by 300 meV.

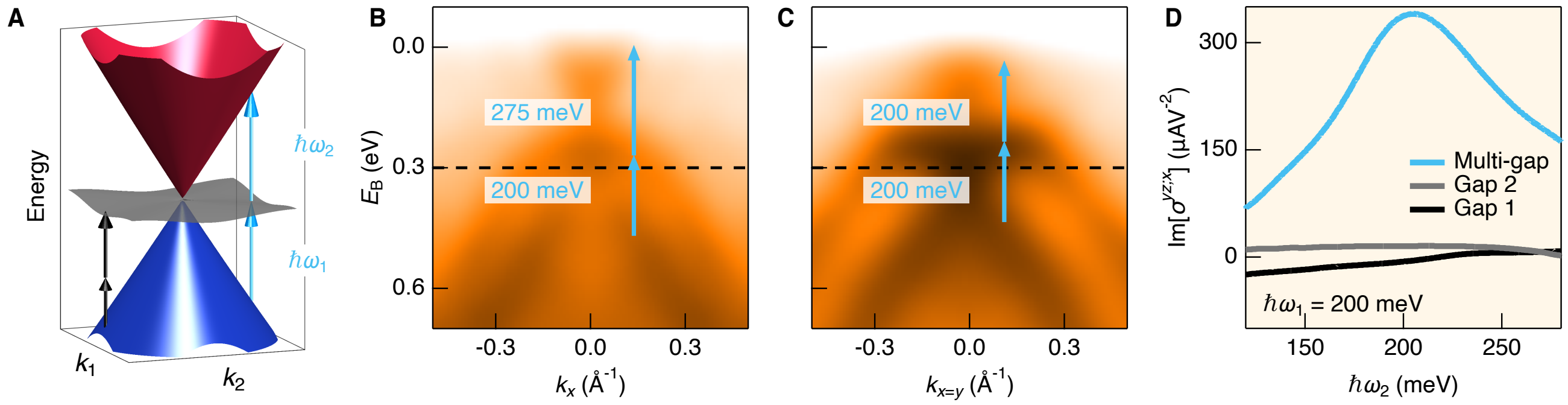

FIG. S4. Giant Sum Frequency Response in Multi-gap topological RhSi. (a), Schematic of second order two-band (black arrows) and three-band (cyan arrows) optical processes for a three-fold chiral fermion. (b), SXARPES along the Γ-X path with dotted line indicating the Fermi level of RhSi. Arrows indicate optical resonances with $\hbar\omega_1 = 200$ meV and $\hbar\omega_2 = 275$ meV. (c), SXARPES along the Γ-M path with dotted line indicating the Fermi level of RhSi. Arrows indicate optical resonances with $\hbar\omega_1 = 200$ meV and $\hbar\omega_2 = 200$ meV.(d), *Ab initio* contributions to the imaginary part of the second-order nonlinear conductivity of RhSi, from two-band processes across gap 1 (black), two-band processes across gap 2 (grey), and multi-gap processes (cyan).

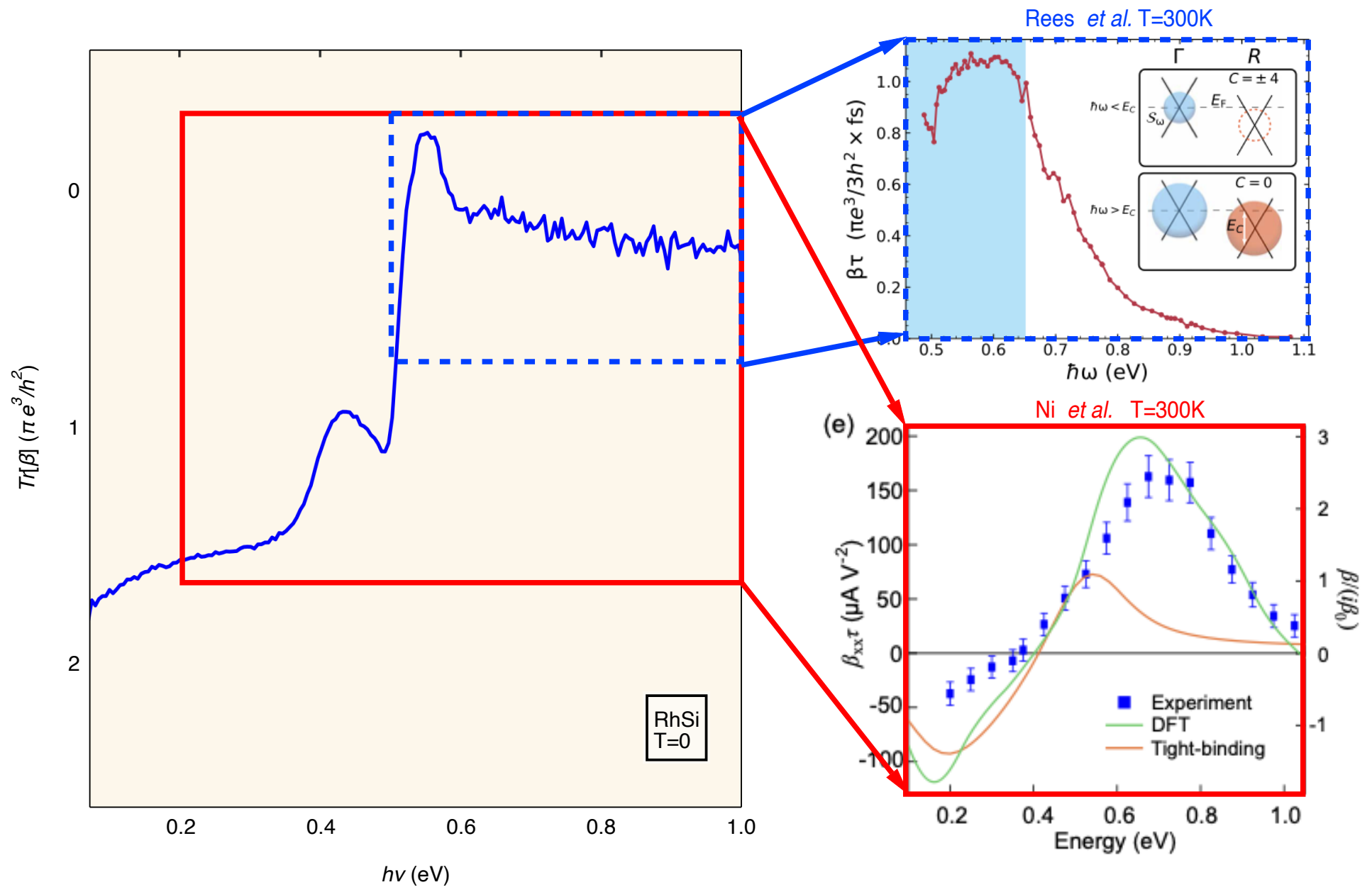

FIG. S5. Previous CPGE experiments do not observe quantized CPGE. Numerical calculation of the CPGE effect in RhSi, with Fermi level determined through SXARPES measurements (left). Experimental THz emission measurements of CPGE from Rees *et al.*, show a spectra reproduced by a spurious non-quantized peak in our calculation (upper right). Experimental THz emission measurements of CPGE from Ni *et al.*, also show a spectra that is consistent with our non-quantized results (lower right). We note that the absolute sign of the $\beta$ tensor is not measurable by THz emission measurements, thus negative photocurrent in our calculations corresponds to positive photocurrent reported by Rees *et al.* [adapted from Rees *et al.* *Sci. Adv.* 6, eaba0509 (2020) and Ni *et al.* *npj Quantum Mater.* 5, 96 (2020)].

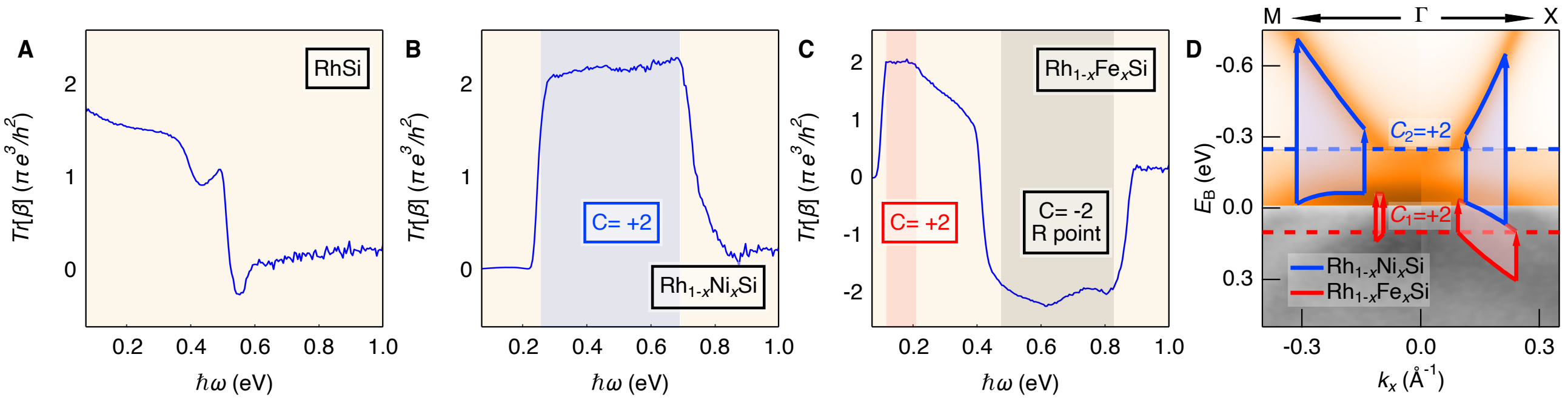

FIG. S6. Quantized CPGE and bulk-photogalvanic correspondence. (a) *Ab initio* CPGE response of RhSi, with native Fermi level determined by SXARPES. (b) *Ab initio* CPGE response of $Rh_{1-x}Ni_xSi$, with Fermi level 250 meV above native RhSi. Blue shaded region indicates a large quantized plateau with $C = +2$. (c) *Ab initio* CPGE response of $Rh_{1-x}Fe_xSi$, with Fermi level 100 meV below native RhSi. Red shaded region indicates a quantized plateau with $C = +2$ and grey shaded region indicates a plateau with $C = -2$. (d) SXARPES measurement on RhSi (grey) and $Rh_{1-x}Ni_xSi$ (gold), and *ab initio* calculation below $E_B = -250$ meV along the M-Γ-X path. Dotted blue/red lines correspond to the Fermi level for $Rh_{1-x}Ni_xSi$/$Rh_{1-x}Fe_xSi$ samples. The blue/red arrows and the shaded regions indicate the optical transitions that give rise to the quantized CPGE plateaus in $Rh_{1-x}Ni_xSi$/$Rh_{1-x}Fe_xSi$ samples.